\documentclass[a4paper,11pt]{article}
\pdfoutput=1 

\usepackage{jcappub} 

\usepackage{rotating}
\usepackage{tablefootnote}

\usepackage{amssymb}
\usepackage{amsmath}

\usepackage{graphicx}
\usepackage{hyperref}

\newcommand\arcmin{\mbox{$^\prime$}}%
\newcommand\arcsec{\mbox{$^{\prime\prime}$}}%

\title{Line-of-sight effects in strong lensing: Putting theory into practice}

\author[]{Simon Birrer,}
\emailAdd{simon.birrer@phys.ethz.ch}
\author[]{Cyril Welschen,}
\emailAdd{cyril.welschen@student.ethz.ch}
\author[]{Adam Amara,}
\emailAdd{adam.amara@phys.ethz.ch}
\author[]{Alexandre Refregier}
\emailAdd{alexandre.refregier@phys.ethz.ch}

\notoc

\affiliation[]{Institute for Astronomy, Department of Physics, ETH Zurich\\ Wolfgang-Pauli-Strasse 27, 8093, Zurich, Switzerland}

\abstract{
We present a simple method to accurately infer line of sight (LOS) integrated lensing effects for galaxy scale strong lens systems through image reconstruction. Our approach enables us to separate weak lensing LOS effects from the main strong lens deflector. We test our method using mock data and show that strong lens systems can be accurate probes of cosmic shear with a precision on the shear terms of $\pm 0.003$ (statistical error) for an HST-like dataset.
We apply our formalism to reconstruct the lens COSMOS 0038+4133 and its LOS. In addition, we estimate the LOS properties with a halo-rendering estimate based on the COSMOS field galaxies and a galaxy-halo connection. The two approaches are independent and complementary in their information content. We find that when estimating the convergence at the strong lens system, performing a joint analysis improves the measure by a factor of two compared to a halo model only analysis.
Furthermore the constraints of the strong lens reconstruction lead to tighter constraints on the halo masses of the LOS galaxies. Joint constraints of multiple strong lens systems may add valuable information to the galaxy-halo connection and may allow independent weak lensing shear measurement calibrations.
}

\begin{document}
\maketitle
\flushbottom



\section{Introduction}

Gravitational lensing is a unique probe for measuring dark matter and dark energy by mapping the mass distribution of the universe on different scales. On the largest scales, weak gravitational lensing surveys measure the linear and non-linear regime of structure formation (see e.g. reviews of \cite{Bartelmann:2001vu, Refregier:2003xq} and references therein). On scales of individual galaxies, time-delay cosmography measures angular diameter distance relations (\cite{Refsdal:1964pi, Kochanek:1996so} or \cite{Treu:2016ul} as a recent review), which gives us information on the background expansion of the universe. On sub-galactic scales, the abundances of dark substructure that can be studied using strong lensing is sensitive to the physical properties of dark matter \cite{Metcalf:2001ig, Dalal:2002tk, Yoo:2006br,Amara:2006gl, Keeton:2009lg, Moustakas:2009ym}.

These different gravitational lensing regimes each developed formalisms to connect their observables with the underlying physical distribution of dark matter being studied. The distinction between these domains effectively reflected the simplifying assumptions they each used. For example, the simplest approach in modeling strongly lensed systems is to describe the process in terms of a single strong perturber and to neglect contributions of other masses along the line-of-sight (LOS). On the other hand, for weak lensing studies, the focus is on integrated tidal distortions due to structure along the LOS, but higher order non-linear stronger lensing effects are typically not included. However, with increasing volumes and quality of data, this distinction between the different regimes is no longer sufficient and an integrated approach to lensing problems needs to be adopted.

There are several examples in strong lensing where the LOS needs to be considered carefully. One is the inference of dark matter substructure properties from Quasar flux ratios \cite{Kochanek:1991xt}, where LOS structure can have a significant impact on this observable and therefore can affect the interpretation of the data \cite{Xu:2012kk, Inoue:2015a, Takahashi:2014yl, Xu:2015va}. Another example is strong lens cosmography \cite{Refsdal:1964pi, Kochanek:1996so, Schechter:1997sw, Oguri:2007fv} since integrated LOS structure in the vicinity of galaxy scale strong lens systems can have a significant impact on relative time-delay measures. These must be taken into account to perform precision cosmographic estimates \cite{Bar-Kana:1996xj, Momcheva:2006oh, Wong:2011wq, Jaroszynski:2014xy, McCully:2016jy}.

Early work studying external shear and ellipticity in gravitational lensing described the impact of LOS mass distribution as an equivalent additional mass sheet at the redshift of the main deflector with uniform surface mass density $\kappa_{\text{ext}}$ \cite{Schneider:1997wt, Keeton:1997zi}. In the literature, the LOS structural parameter $\kappa_{\text{ext}}$ is typically estimated using: (i) imaging and spectroscopy of objects in the neighborhood of the lensing systems \citep[e.g.][]{Keeton:2004fp, Fassnacht:2006dt, Momcheva:2006oh, Anguita:2009ul, Suyu:2010jq, Wong:2011wq, Fassnacht:2011zs}; (ii) weak lensing \citep[e.g.][]{Nakajima:2009fv}; and  (iii) using comparison with cosmological numerical simulations \citep[e.g.][]{Greene:2013wb, Collett:2013bh}. Recently, a general multi-plane lensing framework has been introduced \cite{McCully:2014uf}. The authors later use this approach to test their accuracy in modelling LOS structure using mock position data \cite{McCully:2016jy} for quadrapole lens systems. They point on the need for properly accounting for LOS structure in precision lens modeling.

In this work, we present a set of simplified approximation of the multi-plane framework for accounting for LOS structures. These give reliable reconstructions for strong lens systems around the Einstein ring. The advantage of this is that accuracy is maintained while also allowing us to separate the calculations of the LOS effects from the strong lensing deflections of the main lens. This, in turn, allows us to incorporate LOS modelling into our modelling tools \cite{Birrer:2015yo, Birrer:2016zy} that aim to reconstruct the full extended lens system. We apply our modeling formalism to the lens system COSMOS0038+4133 and demonstrate the power gaining insights into the LOS structure through strong lens image reconstruction.

The paper is structured as follows: In section \ref{sec:multi_plan_lensing} we revisit the geometry of multi-plane gravitational lensing, review the approaches being taken in the literature and introduce our notation. In section \ref{sec:CSB}, we state our approximations, the phenomenological modeling parameterization for strong lens image reconstruction, provide the link to the physical mass distribution in the universe and present test on mock data. In section \ref{sec:cosmos_lens}, we apply our modelling formalism to a strong lens in the COSMOS field. Independently, we perform an environmental analysis based on the galaxies in the vicinity of the lens and show the strength of the combination of strong lens and environment analysis. Finally, in section \ref{sec:conclusion} we draw our conclusions and implication for further work.

Throughout this work, we assume a flat $\Lambda$CDM cosmological model with parameters $\Omega_{\Lambda} = 0.7$, $\Omega_m = 0.3$, $H_0 = 67$ km s$^{-1}$Mpc$^{-1}$.

\section{Multi-plane gravitational lensing} \label{sec:multi_plan_lensing}

In this section, we review multi-plane gravitational lensing, the joint effect caused by multiple lens planes at different distances. We further introduce our notation and state suitable approximations to the full non-linear multi-plane ray-tracing in the regime of one main strong lens. Mathematical aspects of multi plane strong gravitational lensing were studied in \cite{Levine:1993bs, Kayser:1993fq, Petters:1995mk, Petters:1995qx, Petters:2001vl}. Of practical use for our analysis is the multi-plane lens equation \cite{Blandford:1986so, Kovner:1987zz, Schneider:1992vp}.

\begin{figure}[htbp]
\begin{center}
\includegraphics[angle=0, width=120mm]{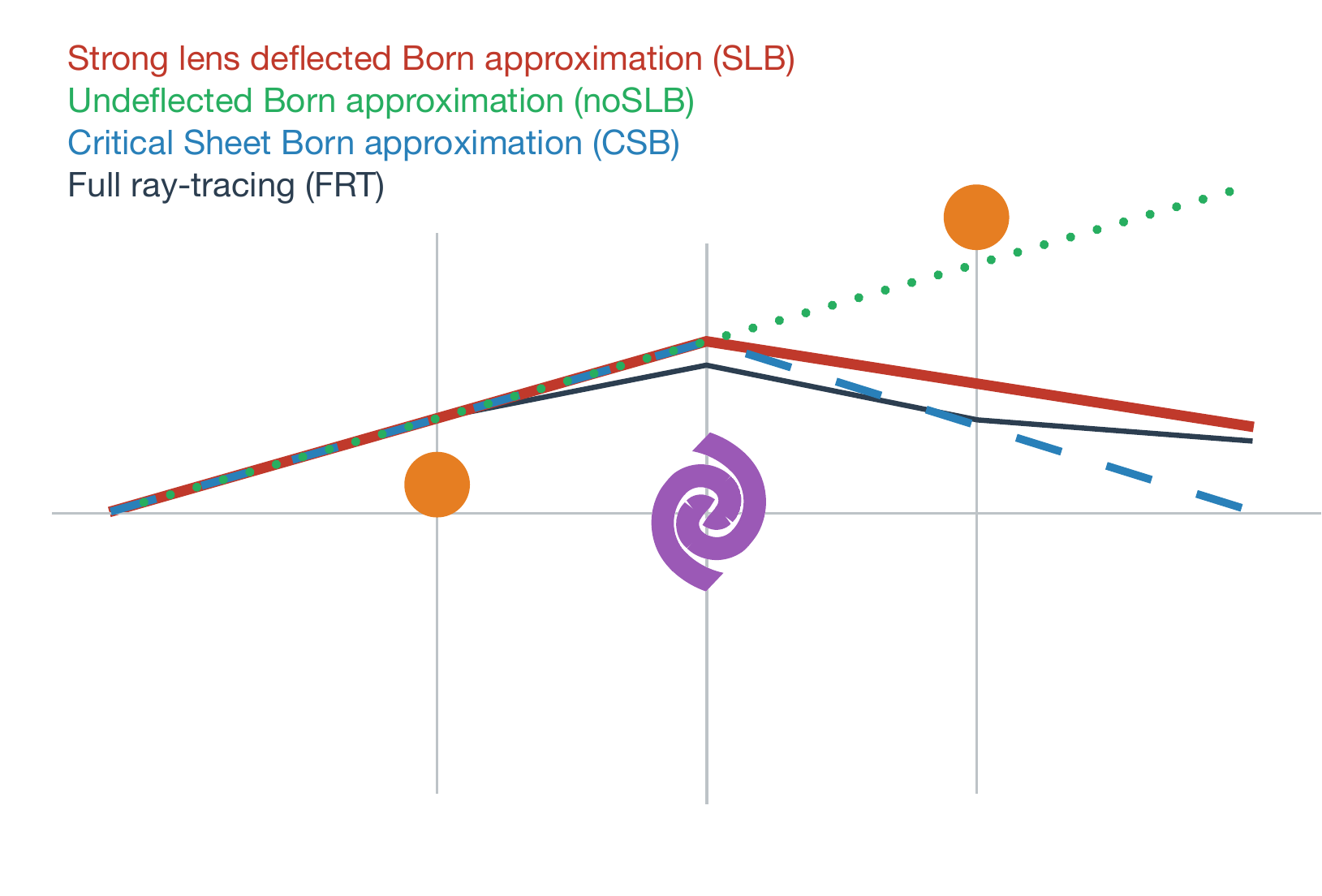}
\caption{Illustration of different approximations of the light path on which to compute the LOS effects. Strong lens deflected Born approximation (SLB, red line): Computation along the strong lens deflected path, which is accurate but leads to non-linear couplings to the strong lens. Undeflected Born approximation (noSLB, green dotted line): Born approximation ignoring the strong lens. This method is inaccurate for accounting of the effects of background perturbers. Critical Sheet Born approximation (CSB, blue dashed line): Replacing the strong lens deflected path (SLB) by a critical mass-sheet deflected path (see section \ref{sec:CSB}). Full ray-tracing (FRT, thin black line): No approximations and every (small) deflector is modeled as a single lens plane.}
\label{fig:light_path_approx}
\end{center}
\end{figure}

\subsection{General description} \label{sec:general_description}
In gravitational lensing, the mapping from source to image is given by the lens equation \citep[e.g. see][for background material]{Narayan:1996yp}
\begin{equation} \label{eqn:lens_equation}
	\vec{\beta}=\vec{\theta}-\vec{\alpha}(\vec{\theta}),
\end{equation}
where $\vec{\theta}$ is the un-lensed angular position, $\vec{\beta}$ is the post lensing position and $\alpha$ is the deflection angle. When studying strong lens systems, the thin lens approximation is widely used. This models the light travel path as straight between lens planes with sharp deflections at the positions of the lenses. In the case of a single lens, the lens equation becomes,
\begin{equation} \label{eqn:thin_lens_approximation}
	\vec{\beta}=\vec{\theta}-\vec{\alpha}(\vec{\theta}) = \vec{\theta} - \frac{D_{ds}}{D_s} \vec{\hat{\alpha}}(D_d\vec{\theta}),
\end{equation}
where $\vec{\hat{\alpha}}(D_d\vec{\theta})$ is the physical bending angle that is linked to $\vec{\alpha}(\vec{\theta})$ through the angular diameter distance  $D_s$ (the angular diameter distance from the observer to the source), $D_d$ (the angular diameter distance from the observer to the lens) and $D_{ds}$ (the angular diameter distance from the lens to the source).

In the case where there are $n$ thin lenses along the LOS, the total mapping is the sum of all the deflections, weighted by their distance relations and evaluated at the light path as
\begin{equation} \label{eqn:multi_thin_lens}
	\vec{\beta}_{s}=\vec{\theta} - \frac{1}{D_s}\sum_{k=1}^n D_{ks} \vec{\hat{\alpha}}_{k}(D_k\vec{\beta}_{k})
\end{equation}
where $\vec{\beta}_{k}$ is the angle under which the $k$'th lens deflects the light ray and $\vec{\beta}_{1} = \vec{\theta}$. The ordering in $D_k$ follows the backwards light path such that the light passes the lens at $k$ before $k-1$ (see also \cite{McCully:2014uf} for a different notation of the same expression).

\subsection{Case of one strong deflectors and several weak ones}
Often in strong lensing, the dominant deflection is due to one single object at a specific redshift with additional deflectors causing secondary weaker effect. This case has been discussed extensively in \cite{Schneider:1997wt}. The black path shown in Figure \ref{fig:light_path_approx} shows an illustration of the light travel path through a multiple lens system (full ray-tracing FRT). One way to simplify the calculation is to treat the smaller additional lenses as tidal perturbers (neglecting higher order terms), i.e.

\begin{equation}
	\vec{\hat{\alpha}}_{k}(\vec{x}) \approx \frac{\partial \vec{\hat{\alpha}}_{k}}{\partial \vec{x}} \vec{x}.
\end{equation}
Higher order effects (flexion) are small for LOS perturber (see e.g. \cite{McCully:2016jy}). The differential form of the lens equation for weak perturbers can be written as

\begin{equation} \label{eqn:perturber_general}
	\frac{\partial \vec{\beta}}{\partial \vec{\theta}}
	= \delta_{ij} - \frac{D_{ks}D_k}{D_s}\frac{\partial \vec{\hat{\alpha}}_{k}}{\partial \vec{x}} \frac{\partial \vec{\beta}_{k}}{\partial \vec{\theta}}.
\end{equation}
This tidal approximation depends on the light path $\vec{x} \equiv D_k\vec{\beta}_{k}$. In general, for a strong lens, the light path $\vec{\beta}_k$ can be highly non-linear for perturbers between source and strong lens ($k > d$).

The undeflected Born approximation (noSLB, green dotted line in Figure \ref{fig:light_path_approx}) computes the light path ignoring any deflector ($\vec{\beta}_{k} = \vec{\theta}$). The noSLB approximation leads to a linear distortion in the lens equation that is given by the distortion matrix ($\Gamma_{ij}$):

\begin{equation} \label{eqn:tidal_distortion}
	\Gamma_{ij}\equiv {\partial \beta _{i} \over \partial \theta _{j}}
	\equiv \left[{\begin{array}{c c }1-\kappa -\gamma _{1}&\gamma _{2}\\\gamma _{2}&1-\kappa +\gamma _{1}\end{array}}\right]
\end{equation}
where $\kappa$ is the convergence and $\gamma_1$, $\gamma_2$ are the shear components of the linear distortion matrix. Another approach, we call it "strong lens deflected Born approximation'' (SLB), takes the main deflector into account when evaluating the effects of the LOS perturbers (the red path shown in Figure \ref{fig:light_path_approx}). The advantage of this approach is that this approximation replicates that of the full ray-tracing calculation with high fidelity relative to noSLB. The problem, however, is that the tidal effect can only be calculated after the light path has been found. This coupling between the ray-tracing and the impact of the secondary lenses makes such a calculation more complex when modelling strong lens systems. Furthermore, a parameterized representation of each individual perturber leads to a high dimensional parameter space whose inference and the covariances between the perturbers and the strong lens is impractical. For this reason, there are significant advantages to finding further simplifications that are able to separate the two computations. The most commonly used method for achieving this is to perform ray-tracing for the main lens (red in Figure \ref{fig:light_path_approx}) with the additional effect from the secondary lenses calculated along the un-lensed path (noSLB, green dotted path of Figure \ref{fig:light_path_approx}).

As we will demonstrate later, ignoring the bending of the main deflector can lead to significant inaccuracies in strong lens calculations. To over come this problem we introduce a new approximation (detailed in the next sections) that better captures effects of secondary perturbers close to strong lenses without a coupling to the ray-tracing by the main lens.

\section{Critical Sheet Born Approximation (CSB)} \label{sec:CSB}
Since most of the information from strong lens systems typically comes from regions close to the Einstein radius $\theta_E$, we focus on finding an approximation that is valid in this region. In particular, we are interested in an approximation of the light path $\beta_k$ that goes through the Einstein radius that is independent of the specific lens model.

It turns out that replacing the actual strong lens by a critical mass sheet provides an accurate description of the light path $\beta_k$ at the Einstein radius and leads to the same LOS effects around the Einstein radius compared to SLB. We call this approximation the Critical Sheet Born (CSB) approximation. In Figure \ref{fig:light_path_approx}, CSB is indicated with the blue dashed line. With such a description of the light path in computing LOS effects, we can avoid non-linear coupling between main deflector and LOS perturbers. In the following, we go through the assumptions and derive the approximations of our approach and provide validity tests based on mock data.

\subsection{Formalism} \label{sec:math_csb}

The distortion effects in equation \ref{eqn:perturber_general} depend on the light path. At the Einstein radius, the light path gets bent such that $\vec{\hat{\alpha}}_{d} = \frac{D_d}{D_{ds}}\theta_E$. A straight path with this one deflection involved is given by
\begin{equation}
 \beta(\theta_E) = 
 \begin{cases}
    \theta_E& \text{if } k < d\\
     \left(1 - \frac{D_{dk}}{D_k} \frac{D_s}{D_{ds}}\right) \theta_E            & k>d.
\end{cases}
\end{equation}

This equation is still dependent on the lens model through $\theta_E$. We adopt a generalized form, that does not explicitly require $\theta_E$ and incorporates the light path above is given by
\begin{equation} \label{eqn:crit_sheet_path}
 \beta \approx 
 \begin{cases}
    \theta& \text{if } k < d\\
     \left(1 - \frac{D_{dk}}{D_k} \frac{D_s}{D_{ds}}\right) \theta            & k>d.
\end{cases}
\end{equation}
This equation describes the solution of the light paths of a critical mass sheet at the strong lens position. Furthermore, it results in a linear description of the distortion effect. In addition to equation \ref{eqn:crit_sheet_path} to compute the distortions, we incorporate the linear distortion effect of the foreground LOS perturbers on the strong lens.
With those stated approximations, the lens equation \ref{eqn:thin_lens_approximation} can be written as
\begin{equation} \label{eqn:combined_approx}
	\vec{\beta}_s = \vec{\theta} - \frac{D_{ds}}{D_s} \vec{\hat{\alpha}}_d(D_d\Gamma^{A}_{ij}\vec{\theta}) 
	 - \left(\Gamma^{B}_{ij} + \Gamma^{C}_{ij} \right) \vec{\theta},
\end{equation}
where $\vec{\hat{\alpha}}_d$ is the physical deflection angle of the main deflector, $\Gamma^{A}_{ij}$ is the distortion matrix at the deflector plane caused by foreground perturbers
\begin{equation} \label{eqn:distortion_lens}
	\Gamma^{A}_{ij} = \delta_{ij} - \sum_{k<d} \frac{D_{k} D_{kd}}{D_d} \frac{\partial \hat{\alpha}^{i}_{k}}{\partial x_j},
\end{equation}
$\Gamma^{B}_{ij}$ is the distortion caused by the same foreground perturbers at the source plane
\begin{equation} \label{eqn:distortion_ol_source}
	\Gamma^{B}_{ij} = \sum_{k<d} \frac{D_{k} D_{ks}}{D_s } \frac{\partial \hat{\alpha}^{i}_{k}}{\partial x_j},
\end{equation}

and $\Gamma^{C}_{ij}$ is the distortion caused by background perturbers on the source plane
\begin{equation} \label{eqn:distortion_ls_einstein_ring}
	\Gamma^{C}_{ij} = \sum_{k>d} \frac{D_{k} D_{ks}}{D_s}\left( 1- \frac{D_{dk}}{D_k} \frac{D_s}{D_{ds}} \right) \frac{\partial \hat{\alpha}^{i}_{k}}{\partial x_j}. 
\end{equation}

The only explicit deflection in equation \ref{eqn:combined_approx} is the main deflector $\hat{\alpha}_d$. As pointed out by \cite{McCully:2016jy}, the non-linear effect of the term $\Gamma^{A}_{ij}$ on $\hat{\alpha}_d$ is important and not taking this effect into account can lead to significant biases in the lens model inference. Furthermore, the LOS structure close to the source plane is of less importance as the light rays are bent and get closer to each other and reduced the induced tidal distortion.

\subsection{Phenomenological parameterization} \label{sec:phenom_csb}
In this section, we discuss what the observables from strong lensing image reconstruction are when the underlying description is approximated by equation (\ref{eqn:combined_approx}). The effect on the lens equation of the LOS structure can be expressed as tidal distortions (equation \ref{eqn:tidal_distortion}). Equation \ref{eqn:combined_approx} becomes

\begin{equation} \label{eqn:framework_true}
	\vec{\beta}_{\text{true}} = \vec{\alpha}_{\text{true}}\left(\left(1-\kappa_d\right) D_\text{d}^{\text{bkgd}}\left[{\begin{array}{c c }1 -\gamma' _{1,d}&\gamma' _{2,d}\\\gamma' _{2,d}&1 +\gamma' _{1,d}\end{array}}\right]\vec{\theta}\right)
	+ \left(1-\kappa_s\right)\left[{\begin{array}{c c }1 -\gamma' _{1,s}&\gamma' _{2,s}\\\gamma' _{2,s}&1 +\gamma' _{1,s}\end{array}}\right] \vec{\theta},
\end{equation}
where $\gamma' = \gamma/(1-\kappa)$ is the reduced shear. $\vec{\beta}_{\text{true}}$ and $\vec{\alpha}_{\text{true}}$  indicate that in the stated form above including the true physical lens model is recovered. $D_\text{d}^{\text{bkgd}}$ states the cosmological background angular diameter distance. Additionally to the main deflector, 6 additional parameters, namely the shear and convergence terms to the lens ($\gamma_{1,d}$, $\gamma_{2,d}$, $\kappa_d$) and the source plane ($\gamma_{1,s}$, $\gamma_{2,s}$, $\kappa_s$) describe the LOS effect.

The convergence terms $\kappa_d$ and $\kappa_s$ lead to particular degeneracies with other lensing effects. A non-zero convergence $\kappa_d$ changes the angular diameter distance according to $D_{\text{d}}^{\text{lens}} = \left(1 - \kappa_d \right) D_{\text{d}}^{\text{bkgd}}$. The angular diameter distance $D_{\text{d}}^{\text{lens}}$ must be considered when computing other physical quantities of the lens, such as lensing potential and kinematics. The angular diameter distance $D_{\text{d}}^{\text{lens}}$ can not be determined from strong lens image reconstruction without relying on other information. The effect of $\kappa_s$ leads to a rescaling of the lens equation (Equation \ref{eqn:lens_equation}, \ref{eqn:combined_approx} or \ref{eqn:framework_true}) without changing image observables. The physical interpretation of the rescaled quantities can change significantly. This is known as the mass-sheet degeneracy \cite{Falco:1985no, Schneider:1995ga, Saha:2000zr, Wucknitz:2002he}.

The convergence effects can be decoupled from the image reconstruction. Instead of modeling $\vec{\alpha}_{\text{true}}$, $\vec{\beta}_{\text{true}}$ $\kappa_d$ and $\kappa_s$ one can model a rescaled lens equation
\begin{equation} \label{eqn:framework_obs}
	\vec{\beta}_{\text{scaled}} = \vec{\alpha}_{\text{scaled}}\left(D_\text{d}^{\text{bkgd}}\left[{\begin{array}{c c }1 -\gamma'' _{1,d}&\gamma'' _{2,d}\\\gamma'' _{2,d}&1 +\gamma'' _{1,d}\end{array}}\right]\vec{\theta}\right)
	+ \left[{\begin{array}{c c }1 -\gamma' _{1,s}&\gamma' _{2,s}\\\gamma' _{2,s}&1 +\gamma' _{1,s}\end{array}}\right] \vec{\theta}.
\end{equation}
The physical interpretation of the inferred deflection angle $\vec{\alpha}_{\text{scaled}}$, source scale $\vec{\beta}_{\text{scaled}}$ and shear terms on the lens plane $\gamma_d''$ change according to the convergences. The actual physical deflection relates to the scaled one as $\vec{\alpha}_{\text{true}} = \left(1-\kappa_d\right)\left(1-\kappa_s\right)\vec{\alpha}_{\text{scaled}}$. The source plane coordinate scales as $\vec{\beta}_{\text{true}} = (1-\kappa_s)\vec{\beta}_{\text{scaled}}$ and the shear induced on the main deflector $\gamma_d''$ scales as
\begin{equation} \label{eqn:reduced_shear_lens}
\gamma''_d = \frac{\gamma_d}{\left(1-\kappa_d\right)^2\left(1-\kappa_s\right)}.
\end{equation}

\subsection{Validity test} \label{sec:validity_approx}
To test the accuracy of the approximations stated in Section \ref{sec:math_csb}, we construct a test scenario and compare the full multi-plane ray-tracing solution with our proposed formalism. For this purpose, we position a singular isothermal sphere (SIS) lens with velocity dispersion $\sigma_v=200$ km s$^{-1}$ at a redshift $z_d = 0.5$ and a source at $z_s=2$. The Einstein radius of this configuration is $\theta_{\text{E}}=0.73"$. We place a single perturber in the form of a Navarro-Frank-White (NFW) profile \cite{Navarro:1997tb} with an angular separation of 8" away from the center of the SIS profile. We chose the mass within a mean over-density of 200$\rho_c$ as $M_{200} = 10^{13}M_{\odot}$. We vary the redshift of the perturber ($z = [0.1,0.3,0.5,0.7,0.9]$) to test our formalism for different LOS positions.

The perturber is a group scale halo close to the main deflector. This is a relatively strong LOS perturber. Any perturber less massive and/or further away in angular separation will have a weaker impact on the deflection angles and will be approximated as well or better in our formalism. The integrated lensing effect from multiple perturbers adds linearly on the shear and convergence terms. The accuracy of multiple perturbers should be valid as long as second order terms are of comparable strength as the single group scale halo. Our formalism is accurate as long as the Born approximation is accurate between observer and main deflector and main deflector and source.

\subsubsection{Convergence maps} \label{sec:convergence_maps}
\begin{figure*}
  \centering
  \includegraphics[angle=0, width=160mm]{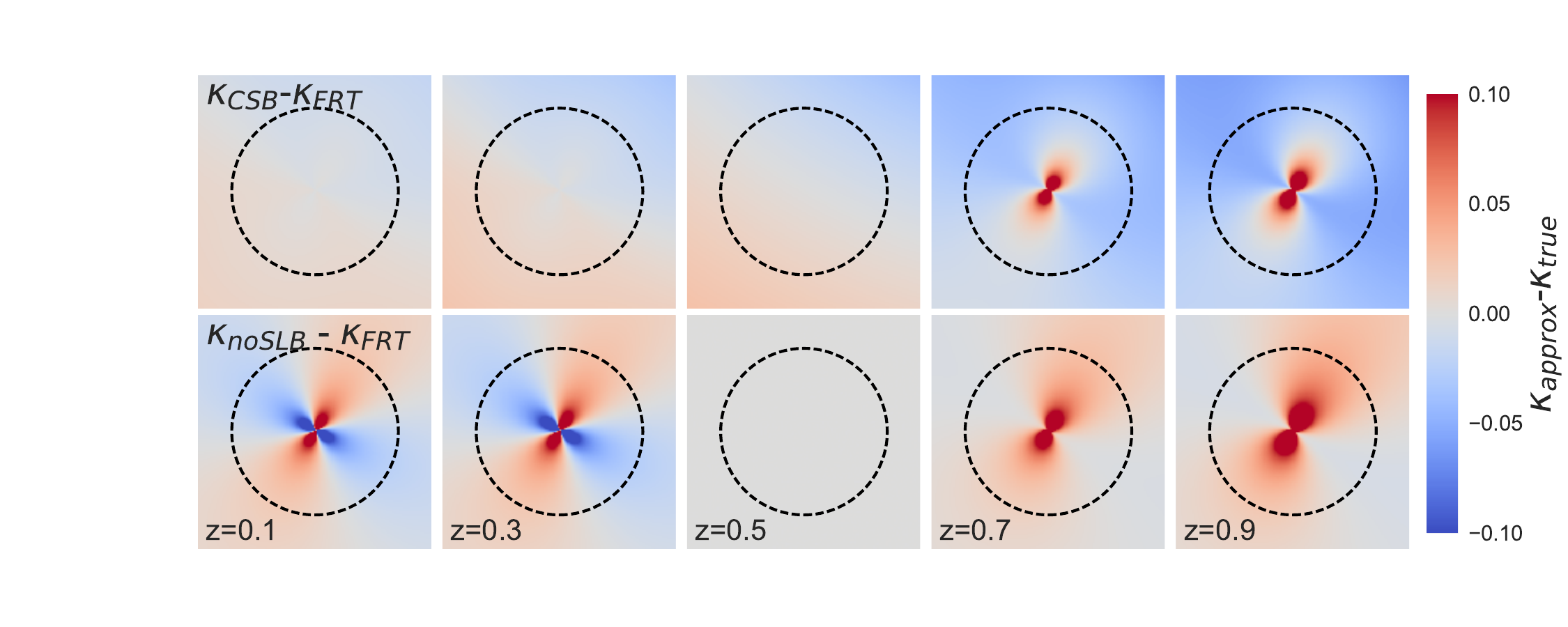}
  \caption{Comparison of different approximations with respect to the convergence prediction. Blue (red) corresponds to an under-(over-)estimation of the convergence by the approximation scheme. Upper row: Relative convergence of CSB compared to FRT. Lower row: Relative convergence of the Born approximation (noSLB) compared to FRT. From left to right: Comparisons with increasing LOS perturber redshift. The main deflector is placed at $z_d = 0.5$. The black dashed circle indicates the Einstein radius of the main deflector.}
\label{fig:environment_z_kappa}
\end{figure*}

We first test the accuracy of the predicted convergence map. In Figure \ref{fig:environment_z_kappa}, we compare the computed convergence maps of the full ray-tracing (FRT) with the approximation of our formalism (CSB). The convergence map for FRT is computed with differential ray-tracing. The top panel of Figure \ref{fig:environment_z_kappa} shows the deviation of the convergence map of CSB to the full solution (FRT) $(\kappa_{CSB} - \kappa_{FRT})$. The lower panel shows a same comparison with the Born approximation (noSLB) of the LOS perturber $(\kappa_{noSLB} - \kappa_{FRT})$. The Einstein radius is plotted in black dashed lines.

The main differences in accuracy occur when the perturber is placed in front of the lens ($z<z_d$). The non-linear effect on the lens model can be well captured by CSB whereas noSLB ignors those effects and leads to significant residuals in the convergence map. The higher order distortion effects of the LOS perturber results in $\Delta \kappa < 0.01$ at the Einstein radius for CSB.

When the perturber is placed between the source and the lens ($z_d>z>z_s$), the two approaches have different predictions but neither of them can predict the convergence map accurately over the entire area of the lensing system to a precision better than $\Delta \kappa \approx 0.1$. The main difference is that CSB reproduces the mean convergence within the Einstein radius while noSLB over-estimates the convergence induced by the perturber significantly. CSB reproduces the mean convergence within the Einstein radius by construction while the induced error in the mean convergence in the noSLB is $\Delta_k \approx 0.05$. This behavior of the two approximations becomes emergent when looking at extended surface brightness simulations (see section \ref{sec:ext_surface_bfightness} below).

\subsubsection{Extended surface brightness} \label{sec:ext_surface_bfightness}
\begin{figure*}
  \centering
  \includegraphics[angle=0, width=160mm]{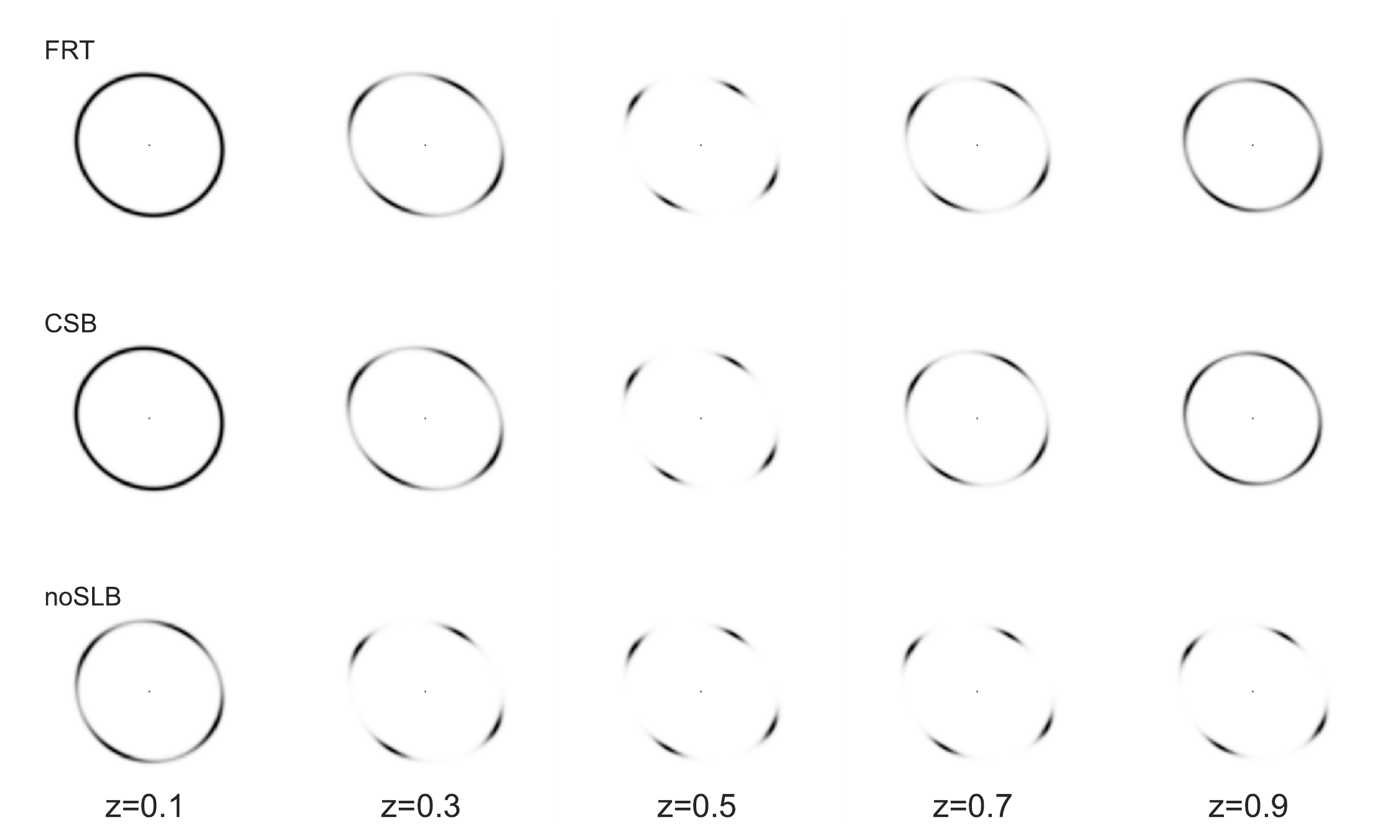}
  \caption{Comparison of different approximations with respect to the image prediction. In this model, a Gaussian source is placed at $z_s=2$ and a main deflector as a SIS profile at $z_d = 0.5$. In addition, a LOS perturber is placed at different redshifts. Top column: Image computed with full ray-tracing (FRT). Middle column: Image computed with the CSB formalism presented in this work. Bottom column: Image computed with the noSLB approximation of the LOS perturber.}
\label{fig:environment_z_image}
\end{figure*}

To test how well extended lensed surface brightness information can be predicted and reproduced by the CSB formalism, we take the same test case of \ref{sec:convergence_maps} and model a Gaussian source surface profile with a width $\sigma = 0.02"$ in the source plane positioned in line with the center of the main deflector. In the absence of external perturbers, this configuration leads to a perfectly circular Einstein ring in the image plane.
In Figure \ref{fig:environment_z_image}, the simulated mock images are shown for the different computations of the LOS structure. For the moment, we do not include any observational effects into the simulation (e.g. PSF and noise). In the top row, the full ray-tracing (FRT) simulations are shown. In the middle row, the predictions of the CSB formalism are shown. We see no distinguishable effect in the image plane. In the bottom row, the predictions with a Born approximation (noSLB) is shown when interpreting the shear terms from the standard weak lensing formalism.
This test shows that the CSB approximation provides a good description for lens and source configurations that form an Einstein ring-like extended structure. We also see that the features in the image are not predicted accurately by a noSLB approximation. In particular, for foreground perturbers (first two columns in Figure \ref{fig:environment_z_image}), the real feature is a sheared Einstein ring/ellipsoid. Lens models that can reproduce elliptical Einstein rings have a non-spherical extended deflection with a point like inner caustic. This requires a particular class of lens models, among which a non-linear sheard spherical lens model is a simple solution to. A SIE, or any other simple elliptical lens models, can not produce an elliptical Einstein ring.

For background perturbers, the CSB approximation for the light paths is valid around the Einstein ring and can accurately predict the observational features. A noSLB approximation overestimates the physically induced tidal distortion and convergence (see also \cite{McCully:2016jy}).

\subsection{Testing the constraining power of LOS inference} \label{sec:inference_test}

We analyze the information content of strong lens imaging data on constraining the LOS structure parameters. To do so, we set up a test case with realistic observational conditions and source surface brightness. We then do a parameter inference and test the recovery of the LOS induced effects.

Specifically, we generate a mock image of a source at redshift $z_s=2$, a main deflector at redshift $z_d = 0.5$ and a LOS perturber at $z_{\text{los}} = 0.1$. The main lens is modeled as a SIS with velocity dispersion $\sigma_v = 200$km s$^{-1}$. The LOS perturber is positioned 8" from the main deflector with an NFW profile with mass $M_{200} = 10^{13.5}M_{\odot}$. We model the extended light emission from the source as a Gaussian light profile with width $\sigma_s = 0.02"$. We compute the observable light emission with sub-pixel resolution ray-tracing, convolution with a HST-like PSF and adding Poisson noise on the observed flux and a Gaussian noise realization of the background comparable to HST image quality of the COSMOS field.

\begin{figure*}
  \centering
  \includegraphics[angle=0, width=160mm]{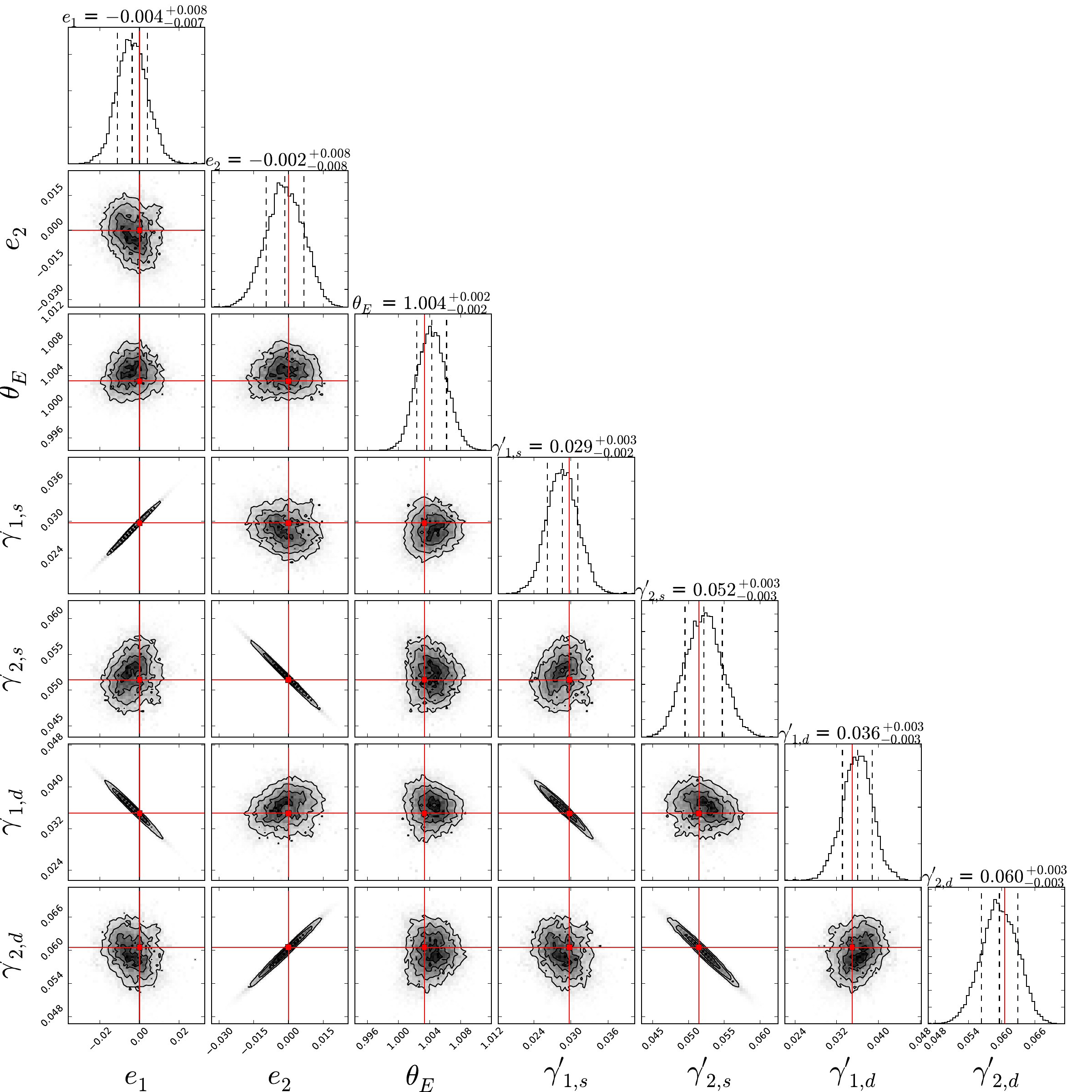}
  \caption{Parameter inference of the mock image described in Section \ref{sec:inference_test}. Red lines indicate the true input for the lens model parameters $e_1$, $e_2$ and $\theta_E$. For the LOS shear parameters, the red lines correspond to the prediction of the CSB formalism modulo the convergence parts (Equation \ref{eqn:framework_obs}, \ref{eqn:reduced_shear_lens}). All parameter posteriors are consistent with the input model. The four external shear terms can be accurately determined to a precision of $\pm 0.003$.}
\label{fig:parameters_mock}
\end{figure*}

We reconstruct the mock imaging data described above to infer the lens model parameter posteriors (including the LOS terms of Equation \ref{eqn:framework_obs}). For the lens model, we choose a smooth power-law elliptical mass profile (SPEMP), which allows for arbitrary elliptical mass distributions and power-law slopes. In the reconstruction modeling, we rescale the source size by $1/(1-\kappa_s)$ to ensure that the same source description is applied in the reconstruction. Source size - power-law slope degeneracies are known and highly depend on the source reconstruction technique applied (see e.g. \cite{Birrer:2016zy}). The more general lens model compared with the mock realization tests more rigorously the capability of recovering the LOS structural parameters. The inference is done with the formalism presented in \cite{Birrer:2015yo} with a Monte Carlo Markov Chain (MCMC), implemented in the \texttt{CosmoHammer} \cite{Akeret:2013nl} software. In Figure \ref{fig:parameters_mock}, the inferred parameter posteriors are illustrated. Red vertical and horizontal lines indicate the input parameters for the lens model and the expected scaled shear parameters caused of the LOS perturber. The inference accurately recovers the expected lens and shear terms provided by the CSB formalism and shows that a separability of main deflector and LOS structure can be made with the given lens model assumptions. The posteriors in the ellipticity of the lens and the external shear terms are degenerate but the effects in the image (i.e. the ellipticity of the ring) can not be fully reproduced by an elliptic lens model configuration of the specifically used parameterization. The marginalized constraints on all the shear terms results in constraints of $\pm 0.003$ (statistical error), a high precision measurement of the reduced shear field at a specific angular position on the sky.

\section{The strong lens COSMOS0038+4133 and its environment} \label{sec:cosmos_lens}
In section \ref{sec:CSB}, we used mock data to show that strong lensing systems can allow us to infer scaled shear terms of the LOS structure with high precision.

The prediction of the same environmental quantities can be independently inferred by specifically modeling the LOS structure. Additionally, an explicit modeling of the mass structure enables us to simultaneously infer the external convergences $\kappa_d$ and $\kappa_s$.

We next apply our formalism to the strong lens system COSMOS 0038+4133 and its environment to test the capabilities of our method on real data. The lens system COSMOS 0038+4133 ($\mbox{R.A.}=10^{\mbox{\scriptsize h}}00^{\mbox{\scriptsize m}}38.2^{\mbox{\scriptsize s}}$ $\mbox{DEC}=+02^\circ 41\arcmin 33\arcsec$ J2000) was chosen as our primary target as the configuration is close to an Einstein ring and there are massive galaxies in its close proximity that potentially add significant external shear and convergence contributions to the lens system. Detailed information about the lens system is provided in appendix \ref{ap:lens_description} and the data, catalogs and derived stellar masses of the galaxies in appendix \ref{ap:cosmos_data}.

First, we perform the strong lens modeling in section \ref{sec:sl_modeling}. Second, we perform the independent LOS structure modeling based on galaxy catalogs in section \ref{sec:cosmos_environment}. Third, we combine the constraints of the two approaches and show the results on the inferred external convergence and on the halo mass of specific galaxies in the vicinity in section \ref{sec:cosmos_combined}.

The external convergences are important for many strong lens studies that involve the knowledge of the physical scales at the lens and/or source plane. The mass-sheet degeneracy prevents one from constraining the convergence terms from the strong lens modeling alone. An independent LOS structure modeling predicts both, shear and convergence. The combined constraints on the shear terms form strong lens modeling and LOS reconstruction enables constrain the convergence terms of the LOS modeling more precisely.

\subsection{Strong lens reconstruction} \label{sec:sl_modeling}

We model a 120$^2$ pixels cutout centered on the lensing galaxy. The lensing galaxy light profile is modeled with an elliptical S\'ersic profile \cite{Sersic:1968rm}. The source is modeled with shapelet basis sets \cite{Refregier:2003eg} with $n_{\text{max}} = 10$, which corresponds to 66 basis functions. The shapelet scale is chosen to be $\beta = 0.016"$, which provides a good fit to the data (modulo mass-sheet transform, see e.g. \cite{Birrer:2016zy}). For the lens model, we model a Singular Isothermal Ellipsoid (SIE) and in addition the external reduced shear components of Equation (\ref{eqn:framework_obs}). We use the framework of \cite{Birrer:2015yo} as in Section \ref{sec:inference_test} to infer the parameter posteriors. In this particular inference, we further assume that the lens mass of the SIE is centered at the position of the luminous profile of the lensing galaxy.

\begin{figure*}
  \centering
  \includegraphics[angle=0, width=160mm]{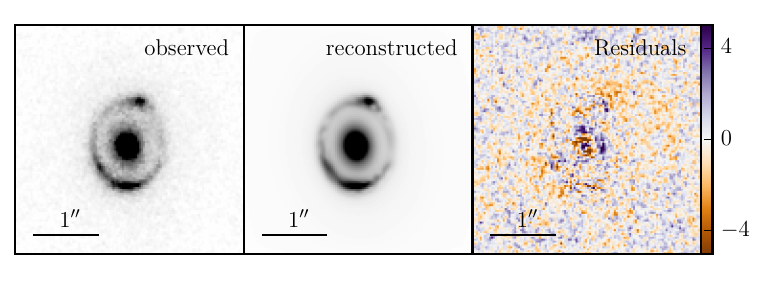}
  \caption{The strong lens COSMOS 0038+4133. In the left panel, the observed HST F814W image is illustrated. In the middle panel, the best fit reconstruction is shown. The right panel shows the reduced residuals. The model allows to reconstruct the arc features in detail. The largest residuals are present in the central part of the lens galaxy.}
\label{fig:COSMOS0038_4133}
\end{figure*}

Figure \ref{fig:COSMOS0038_4133} shows the original HST F814W image (left), the best fit reconstructed model (middle) and the reduced residuals (right). In Table \ref{tab:parameter_constraints} the lens model parameter posteriors are stated for COSMOS0038+4133. We see that the precision on the shear terms is comparable to the mock example in Section \ref{sec:inference_test} (Figure \ref{fig:parameters_mock}). With the stated model assumptions, the shear parameters associated with the LOS structure can be inferred with an uncertainty of $\pm 0.002$. 

\begin{table}
\centering 
\begin{tabular}{ c | c | c |} 
&max likelihood & 1-$\sigma$ posteriors\\
\hline \hline
$\theta_E$& 0.663&$0.663\pm0.001$\\
$e_1$ &-0.105&$-0.104\pm0.006$\\
$e_2$ & 0.018&$0.017\pm0.004$ \\

$\gamma_1^d$ & -0.114&$-0.115\pm0.002$\\
$\gamma_2^d$ & -0.061& $-0.061\pm0.002$\\
$\gamma_1^s$ & -0.219&$-0.218\pm0.003$\\
$\gamma_2^s$& -0.034&$-0.035\pm0.002$\\

\end{tabular}
\caption{Lens model parameter inference for COSMOS 0038+4133. Not included in this list are the lens light model parameters. The column labeled "max likelihood'' shows the parameter position of the global maximum in the likelihood. The image reconstruction of this lens model is shown in Figure \ref{fig:COSMOS0038_4133} middle panel. The second column shows the 1-$\sigma$ marginal posteriors.}
\label{tab:parameter_constraints} 
\end{table}

\subsection{Halo rendering} \label{sec:cosmos_environment}

We estimate the mass distribution in the vicinity of COSMOS 0038+4133 by linking the galaxies in the catalog (see appendix \ref{ap:cosmos_data} for details about the galaxy catalog used) to the underling matter distribution. Similar approaches have been taken by \citep[e.g.][]{Anguita:2009ul, Faure:2011jx} on the COSMOS field. Direct halo mass measurements come for example from galaxy-galaxy lensing \citep[e.g][]{Brainerd:1996bu, Leauthaud:2010mx}. Indirect methods use galaxy clustering \citep[e.g][]{Seljak:2000cz, Zehavi:2002yt} or abundance matching \citep[e.g][]{Kravtsov:2004go, Guo:2010rj, Behroozi:2010yr}. Phenomenological evolutionary models incorporating galaxy evolution in dark matter halos are presented in \citep[e.g.][]{Behroozi:2013iu, Birrer:2014em}.

We use the stellar-to-halo mass relation (SHMR) by \cite{Leauthaud:2012qj}, which is based on simultaneously modeling galaxy-galaxy lensing, galaxy clustering and abundance matching on data from the COSMOS survey. The scatter in the SHMR is described as a log-normal probability distribution function $M_* = f_{\mbox{\tiny SHMR}}(M_h)$ and its inverse \cite{Behroozi:2010yr}. We use the best fit parameters found in three redshift bins, which can be found in \cite[][Table 5]{Leauthaud:2012qj}. For $z>1$ we use the same parameters as for $z\in [0.74, 1]$.

Uncertainties in the involved stellar mass estimates propagate non-linearly through the SHMR and affect the halo mass function, in particular it leads to a more frequent sampling of rare high mass halos. This is in contradiction to the method applied to determine the SHMR, which is based on a given fixed halo mass function. To avoid this inconsistency, we apply a conditional rendering on a fixed halo mass function.

For the spacial distribution of the mass, we assume spherical symmetric Navarro-Frenk-White (NFW) profiles \cite{Navarro:1996ko}. The masses inferred are taken to be the masses enclosed in a mean over-density of 200$\rho_{\text{crit}}$. The mass and redshift dependence of the NFW concentration parameter $c$ is taken from \cite{Neto:2007fu, Ludlow:2014fs}. The object-by-object dispersion in $c$ at fixed halo mass and redshift is assumed to be log-normal as 0.08 dex. The lensing distortions of the NFW profiles are computed following e.g. \cite{Bartelmann:1996an, Wright:2000og}. Uncertainties in the measurements and modeling (i.e. stellar mass, SHMR, mass-concentration relation, redshift) can be incorporated by rendering different realizations of the uncertain quantities and propagate their uncertainties through their dependencies.

We only model over-dense regions of the universe explicitly. This leads to a manifestly over-dense universe compared to the assumed underlying cosmological model. \cite{McCully:2016jy} compensated this effect by ray-tracing through a homogeneous under-dense universe populated with over-dense halos. We chose a different approach. The necessary and sufficient requirement to keep the mean curvature of the universe to the one imposed by the background is that the mean convergence of all angular directions in the universe to all redshifts is zero $\left<\kappa\right> = 0$. A homogeneous under-dense mass distribution contributes a negative convergence $\kappa_{m<0} < 0$. The model thus has to satisfy

\begin{equation}
0 = \left<\kappa_{\text{halo}} + \kappa_{m<0}\right> = \left<\kappa_{\text{halo}}\right> + \kappa_{m<0}.
\end{equation}
This results in a shift of the convergence estimate of
\begin{equation}
\kappa_{\text{render}} = \kappa_{\text{halo}} - \left<\kappa_{\text{halo}}\right>.
\end{equation}
The term $\left<\kappa_{\text{halo}}\right>$ is the mean convergence in a randomly sampled distribution of the galaxies in the field. This method is valid when the universe is homogeneous on the scale being rendered.

\begin{figure*}
  \centering
  \includegraphics[angle=0, width=140mm]{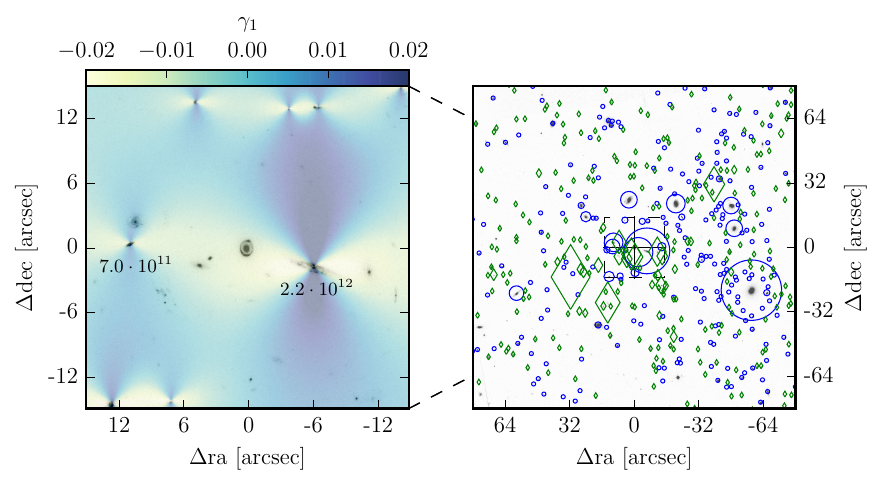}
  \caption{Illustration of the environment of the lens COSMOS0038+4133 and its influence on the shear at the position of the strong lens. Left panel: 30" cutout of the HST COSMOS field centered around the lens system. The shear contribution on the source plane $\gamma_{1,s}$ of the neighboring galaxies is shown. Right panel: 160" zoom-out of the left panel. Blue circles indicate galaxies between the observer and the strong lens. Green diamonds indicate galaxies between the strong lens and redshift $z=2.7$. The size of the circle/diamond indicates the shear strength induced on the strong lens system from the galaxy.}
\label{fig:shear_field}
\end{figure*}

Figure \ref{fig:shear_field} illustrates the environment of COSMOS0038+4133 and the influence on the shear of the nearby galaxies in two different zoom-out regions. The shear estimate converges with a mask of 6.5 arcmin around the strong lens system. For the final sampling, we take a mask of 13 arcmin around the strong lens system. In the selected area, more than 22'000 galaxies are found in the catalog and the contribution to the shear and convergence of each of them is rendered individually and summed up according to equation \ref{eqn:distortion_lens}, \ref{eqn:distortion_ol_source} and \ref{eqn:distortion_ls_einstein_ring}.

The conservative stellar mass estimate uncertainties of the galaxies and the uncertain SHMR, especially at high stellar masses, results in weak constraints on the shear and convergence estimates. Nevertheless, a clear direction (sign) of the shear components is inferred (see Figure \ref{fig:shear_comb} and further discussions in section \ref{sec:cosmos_combined}). Furthermore, the mass rendering indicates a highly over-dense LOS, which is not surprising given the way we selected the lens system.

\subsection{Combining mass rendering and strong lens inference} \label{sec:cosmos_combined}
\begin{figure*}
  \centering
  \includegraphics[angle=0, width=140mm]{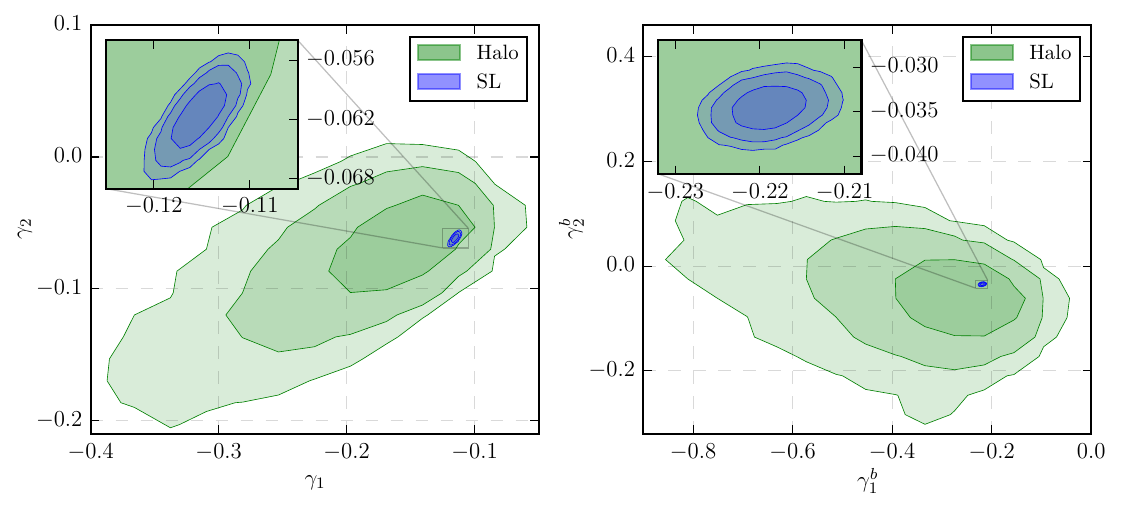}
  \caption{Reduced shear at the lens position $\gamma_{1,2}$ (left panel) and the source position $\gamma_{1,2}^b$ (right panel). Green contours: The 1-2-3 $\sigma$ posteriors of the halo rendering approach based on the galaxy catalogue (section \ref{sec:cosmos_environment}). Blue contours: The 1-2-3$\sigma$ posteriors of the strong lens image reconstruction based on HST image (section \ref{sec:sl_modeling}). Both independent approaches are in agreement with each other. The strong lensing analysis provides much tighter constraints on the reduced shear components. In addition, the mass rendering approach simultaneously provides information about the external convergences and the halo masses of each individual galaxy in the catalogue.}
\label{fig:shear_comb}
\end{figure*}

Figure \ref{fig:shear_comb} shows the posterior distributions of the scaled reduced shear components. We see a consistent inference of the two independent methods in all the four shear terms. The strong lens image analysis leads to much tighter constraints on the shear terms as compared to the halo rendering approach.

From the halo rendering, we can compute the probability distribution
\begin{equation}
 P_{\text{halo}}(\gamma_{1,d},\gamma_{2,d}, \kappa_d, \gamma_{1,s},\gamma_{2,s} \kappa_s, M_h^1, c^1, ..., M_h^n,c^n),
\end{equation}
which involves all the galaxies (mass and profile parameters) and the lensing quantities, including the external convergence terms. The strong lens modeling provides the probability $P_{\text{SL}}(\gamma''_{1,d},\gamma''_{2,d}, \gamma'_{1,s},\gamma'_{2,s})$. For the combined analysis, the two probability distributions can be taken as two independent unnormalized likelihoods as

\begin{equation}
	P_{\text{Halo+SL}} \propto P_{\text{Halo}} \cdot P_{\text{SL}}.
\end{equation}
The two probabilities $P_{\text{Halo}}$ and $P_{\text{SL}}$ are represented by a discrete sample. To numerically combine the two likelihoods, we use kernel density estimators for $P_{\text{SL}}$ to evaluate for each sample in $P_{\text{Halo}}$ a probability weight from the strong lens analysis. The marginalized errors on the parameters of interest come from the samples of $P_{\text{Halo}}$ with their weights from $P_{\text{SL}}$. This is a simple Monte Carlo approach \citep[see e.g.][for an other application in astronomy]{Busha:2011wh}.

\begin{figure*}
  \centering
  \includegraphics[angle=0, width=140mm]{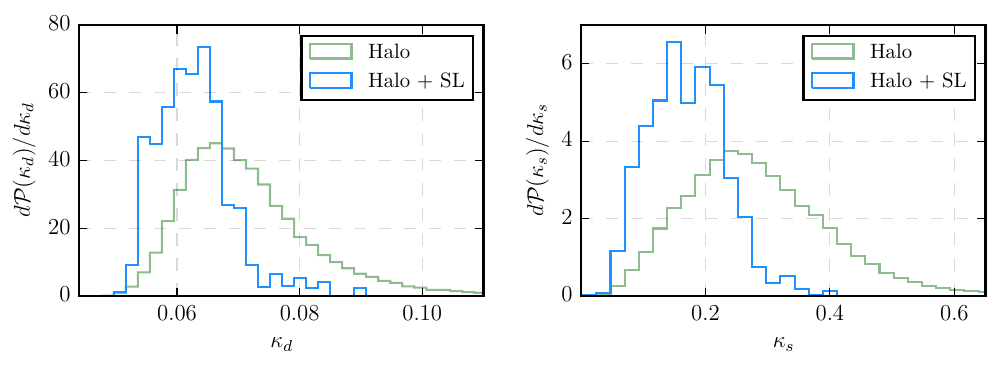}
  \caption{The inferred external convergence at the lens plane $\kappa_d$ (left) and the source plane $\kappa_s$ (right). The marginalized posteriors are shown for the halo rendering only constraints (green) and with the additional constraints on the shear from the strong lens modeling (blue). The tight constraints on the reduced shears of the strong lens analysis leads to double the precision in the inferred convergences.}
\label{fig:convergence_combined}
\end{figure*}

Figure \ref{fig:convergence_combined} shows the inferred external convergence at the lens plane $\kappa_d$ (left) and the source plane $\kappa_s$ (right). The halo rendering only constraints are drawn in green and halo rendering and strong lens reconstruction joint constraints are drawn in blue. The tight constraints on the reduced shears of the strong lens image reconstruction leads to a significant increase in precision of the inferred convergence values.

\begin{figure}
  \centering
  \includegraphics[angle=0, width=80mm]{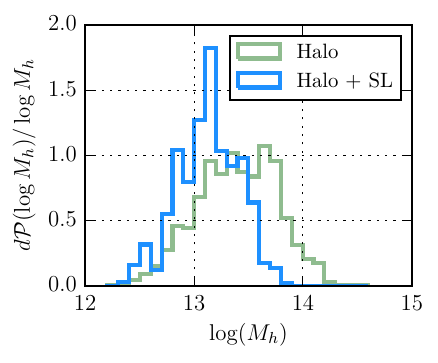}
  \caption{Halo mass constraints for the galaxy 62 arcsec away along the LOS of the strong lens system at $z=0.342$. This is the galaxy with the large blue circle in the lower right half of Figure \ref{fig:shear_field}, right side.}
\label{fig:halo_mass}
\end{figure}

The additional constraints on the scaled shear terms from the strong lensing image reconstruction can also help constrain the halo masses of individual galaxies neighboring the strong lens system. Figure \ref{fig:halo_mass} shows the constraints on the halo mass for a selected massive and nearby galaxy. The strong lens inference implies for this particular galaxy that a very high halo mass can be ruled out. In particular for a nearby massive galaxy, including the strong lens information, the posterior on the halo mass shifted by 0.4 dex to lower halo masses. The statistics of one single strong lens system does not allow to draw significant constraints on the SHMR. Joint constraints of multiple strong lens systems may add valuable information to the galaxy-halo connection.

\section{Conclusions} \label{sec:conclusion}

We have presented a method to infer line of sight integrated lensing effects for galaxy scale strong lens systems through image reconstruction. Our approach enables us to separate weak lensing line of sight effects from the main strong lens deflector and allows a physical interpretation of both effects in parallel without relying on additional estimates in the image reconstruction.
In particular, our approach reconstructs the observed shape and structure of extended arc and ring structure in strong lens systems and the approximations of the light paths and the parameterization allows us to decompose non-linear shear effects caused by intervening structure from the main deflector. We validated our method on mock data and demonstrated that strong lens systems can be accurate and precise probes of cosmic shear. In a second step, we applied our formalism to the COSMOS field. We reconstruct the HST image, including the extended strong lens features. Independently, we modeled the LOS structure inferred from halo rendering using galaxy position, redshift and stellar mass estimates. When performing a combined analysis of our formalism with the halo rendering approach, we improve the constraints on the external convergence by a factor of two compared with a halo-rendering only analysis of the environment.

Strong lenses also allow a very precise direct shear measurement at few specific positions on the sky. This is complementary to galaxy shape weak lensing measurements. Including strong lensing constraints in large scale lensing surveys might thus help in calibrating galaxy shear measurements and constraining the mass distribution in the universe. Furthermore with increasing samples of strong lenses, one can gain insights into the galaxy-halo connection by combining strong lens image modeling and halo rendering of their environments.

\section*{Acknowledgments}
We acknowledge the import, partial use or inspiration of the following python packages: CosmoHammer \citep[][]{Akeret:2013tg}, FASTELL \citep[][]{Barkana:1998sk}, numpy \footnote{www.numpy.org}, scipy \footnote{www.scipy.org}, astropy \footnote{www.astropy.org}. This work has been supported by the Swiss National Science Foundation (grant 200021\_149442/1 and 200021\_143906/1). This research has made use of the NASA/ IPAC Infrared Science Archive, which is operated by the Jet Propulsion Laboratory, California Institute of Technology, under contract with the National Aeronautics and Space Administration.

\appendix

\section{The strong lens COSMOS 0038+4133} \label{ap:lens_description}

The strong lens COSMOS0038+4133 was discovered and first quantified by \cite{Faure:2008gd}. This system has a lensing arc including four images of a source object, see Figure \ref{fig:COSMOS0038_4133} left panel. The redshift of the lens in \cite{Faure:2008gd} was calculated with the publicly available \textit{Le Phare} photometric redshift estimation code using 8 bands, to be $z=0.89^{+0.05}_{-0.03}$ at 68\% confidence level. \cite{Ilbert:2009at} released a revisited redshift estimate inferred from 30 bands at $z=0.733^{+0.008}_{-0.012}$ at 68\% confidence level. In this analysis we take the more recent redshift estimate of the lens. The Einstein radius is about $\theta_E = 0.73"$ and the effective radius of the lens galaxy $R_{\text{eff}} = 0.72"$. The magnitude of the lensing galaxy was determined mag$(I_{814w}) = 20.4$ and the maximum brightness of the ring as mag$"^{-2}(I_{814w}) = 20.5$. The (unknown) redshift of the source was placed to be at twice the co-moving distance to the lens at $z_s = 2.7$ for their lens kinematics and mass estimates. This choice maximizes the lensing efficiency and therefore provides lower bounds on the mass of the lensing galaxy. We adopt the same choice in our analysis for the source redshift.

\section{COSMOS data and catalogues} \label{ap:cosmos_data}

The COSMOS field (see e.g. \cite{Scoville:2007rq}) has been continuous covered by the HST Advanced Camera for Surveys (ACS) Wide Field Channel (WFC) in filter F814W. The median exposure depth is $2028\,\mbox{s}$ and the limiting point-source depth is F814W$_{AB}=27.2\,(5\sigma)$. This results in a 50\% completeness for galaxies with a radius of $0.25\arcsec$ at $I_{AB}=26.0$ mag. The images were combined with the MultiDrizzle software \cite{Jedrzejewski:2005ul} where the final resolution of the drizzled data is $0.03\arcsec/\mbox{pixel}$. We use the third public release v2.0 of the COSMOS ACS data (31. Oct. 2011) \footnote{STScI-MAST: \url{http://archive.stsci.edu/} or IPAC/IRSA: \url{http://irsa.ipac.caltech.edu/data/COSMOS/}}. Details of HST ACS/WFC observations, the data calibration and processing procedures are explained in \cite{Koekemoer:2007hf}. The raw data of the ACS WFC were corrected for the charged transfer inefficiency by \cite{Massey:2010dg}.

The COSMOS field provides, apart from the HST coverage, a wealth of additional data products to reconstruct the environment of the lens. Detailed information of the HST observations can be found in \cite{Scoville:2007rq}. We take the redshifts and magnitudes from the COSMOS photometric redshift release \cite{Ilbert:2009at} for the neighboring galaxies, including apparent magnitudes provided by \cite{Capak:2007jq}.

The photometric redshifts in \cite{Ilbert:2009at} were calculated using fluxes in 30 different bands (broad and narrow bands covering UV, visible near-IR and mid-IR). Up to $z\sim2$ the accuracy is $\sigma_{\Delta z/(1+z_s)}=0.06$ at $i_{\mbox{\tiny AB}}^+ \sim 24$, where $\Delta z = z_s - z_p$ and $z_s$ are the spectroscopic redshifts of a comparison sample. We do not include the redshift uncertainty in our analysis as they are of order the cosmological uncertainties. 

We use the NIR $K$ band to calculate luminosities and the broad bands $g^+$ and $i^+$ as a color indicator to estimate the mass-to-light ratios. We take the color dependent mass-to-light ratio by \cite{Bell:2001jz, Bernardi:2003ok} in the functional form of \citep[e.g.][]{Bell:2003bh}

\begin{equation}\label{eq:masstolightratio}
\log\left(\frac{M}{L_\mathcal{B}}\right) = a_\mathcal{B}^\mathcal{C} + b_\mathcal{B}^\mathcal{C} \cdot \mathcal{C}\;,
\end{equation}
where in our case $\mathcal{C}=(g^+ - i^+)$ and $\mathcal{B}=K$. To calibrate the coefficients $a_\mathcal{B}$ and $b_\mathcal{B}$ we use a sample of stellar masses from the COSMOS group membership catalog \citep{George:2011wb} where the stellar masses are calculated according to the method described in \cite{Leauthaud:2010mx}. We split the calibration sample in six redshift bins and infer the coefficients for the different redshift samples independently.

\bibliography{BibdeskLib}

\end{document}